\documentclass{article}

\usepackage{arxiv}
\usepackage[utf8]{inputenc}

\usepackage{longtable}
\usepackage{authblk}
\usepackage{graphicx}
\usepackage{amssymb}
\usepackage{amsmath}
\usepackage{amsthm}
\usepackage[noend]{algpseudocode}
\usepackage{algorithm}
\usepackage{titlesec}
\usepackage[utf8]{inputenc}
\usepackage{enumerate}
\usepackage{subcaption}
\usepackage{graphicx}
\usepackage{nicefrac}
\usepackage{microtype}

\makeatletter
\def\BState{\State\hskip-\ALG@thistlm}
\makeatother

\title{Progress Metrics in DAG-based Consensus}
\author{Quan Nguyen, James Henderson, Egor Lysenko}
\affil{FANTOM}

\begin{document}
	\maketitle
	
\begin{abstract}
Lachesis protocol~\cite{lachesis2021} leverages a DAG of events to allow nodes to reach fast consensus of events. This work introduces DAG progress metrics to drive the nodes to emit new events more effectively. With these metrics, nodes can select event timing and can choose previous events as parents for their own new events. Our results show that our event timing and parent selection methods can help reaching consensus quicker and thus can reduce lower time to finality significantly.	
\end{abstract}
	

\section{Introduction}\label{SEC:Intro}

Lachesis protocol~\cite{lachesis2021, stairdag} is a BFT consensus protocol that reduces communications overhead as events can be reused in the form of event DAG. 
Nodes participating in Lachesis consensus produce events asynchronously. Nodes generate event blocks to form a DAG of events, each of which contains transactions. Each event block has references to parent event blocks. 
For an event block $e$ and it's parents $e_p$, there is no restriction about the creator nodes of $e$ and that of $e_p$.
Each node computes Roots to find Atroposes in order to reach consensus on finalized event blocks.

Each node can generate new own event block and can choose which previous events as the parents for this new event. Throughout this paper, the terms `event block` and `event` are used interchangeably. Event generation involves two steps \emph{Event Timing} and \emph{Parent Selection}.
\begin{itemize}
\item Event Timing: node can decide by itself when it can generate a new own event block.
\item Parent Selection: this is the process of select a suitable set of previous event blocks and use them as parents (aka. $e_p$) for the new event.
\end{itemize}

For the network to perform efficiently, nodes need a set of criteria so as to produce a new event in an effective manner. If nodes produced too many events, it could increase communication overheads as peers will need to handle those events. At the same time, many of such events are possibly abundant in the sense that they do not necessarily contribute to reach a new root. Thus, it is important to know whether a node should emit a new event and whether they should choose certain events as the parents for this new event.

In this work, we introduce two new metrics, which measure the DAG progress of an event and can be used to optimize the way that a node emits a new event. The two metrics are \emph{Quorum Indexer} (QI) metric and \emph{Root Knowledge} (RK) metric. The metrics enable nodes to determine whether a proposed new event block will have a positive contribution towards reaching a new root.

Based on the two metrics, we have defined new criteria for event timing and parent selection for nodes.
In event timing, node can decide itself whether creating a new event will be beneficial. Leveraging the metrics ensures that nodes attempt to create events only if they can produce substantial DAG progress toward finality.
The metrics can also improve parent selection, in which nodes can select previous events as candidate parents for its new event, and thus it can improve the contribution of each event toward reaching finality.

We compare the two metrics in a set of experiments. Our experimental results show that RK metric outperformed QI metric.
Our results also show that event timing and parent selection based on the RK metric have significantly reduced the node overheads (i.e. avoid generate excessive events and handle them) and time-to-finality (TTF).  Efficiency is achieved as nodes will create events when they are able to make substantial DAG progress, reducing the computational time and resources used for processing and transmitting events.

\section{DAG Progress Metrics}

This section describes two new DAG progress metrics. These metrics are used in event timing and parent selection.

\subsection{Quorum Indexer Metric}\label{sec:QImetric}
Quorum Indexer (QI) is a measure of DAG progress. QI metric is based on the sequence numbers of HighestBefore events.

Let $v$ be a node. For a local DAG of $v$, we determine each highest (latest) event $e_i$ for each node $i$. HighestBefore calculations are performed for each node $v$ as follows.
\begin{itemize}
\item Median: Each node $v$ computes the highest events $e_i$ for each node $i$ in a local DAG of $v$. For each $e_i$, it then finds the sequence number of $e_i$ in the subgraph. Then it calculates the median sequence number weighted by the stake of each highest event creator's stake. Intuitively, this gives an estimate of the median highest event created by $i$ in all nodes. 
\item Current Self: A node $v$ finds the sequence number of highest event created by $v$ in the subgraph of its own.
\item New Self: A node pre-computes the sequence number of highest event created by $v$ that would be in the subgraph if it created such a new event.
\end{itemize}

The above three sub-metrics for node $v$ are combined into a single QI metric of DAG progress produced by a new event. To do so, the three metrics are compared and transformed using a piecewise linear function into a metric in the range $[0,w_i/W]$, where $w_i$ is the stake of node $v$ and $W$ is the total stake of all nodes. 

The QI metric $h^{(e)}$ is defined as summation across all nodes $i$ to give a final metric of DAG progress for a new event $e$. The metric is in the range [0,1], with 0 indicating little DAG progress, and 1 indicating significant DAG progress.

\subsection{Root Knowledge Metric}\label{sec:RKmetric}
Progress in Lachesis consensus is achieved via the production of new roots and frames. Root Knowledge (RK) metric is a more effective metric as it is built on the knowledge of previous roots.

An event $e$ is a root when in $e$'s subgraph quorum roots of the previous frame forkless-cause $e$. This means that within $e$'s subgraph quorum roots of the previous frame are each in the subgraph of events created by quorum nodes, potentially with different sets of quorum nodes for each previous root. Further, there can be no forks detected for any of the nodes involved in confirming the forkless-cause condition.

\subsubsection{Notations}
The progress of an event $e$ toward meeting the required conditions for the next root can be described via a root knowledge matrix $\mathbf{K}^{(e)}$ (or simply $\mathbf{K}$).

We write knowledge of previous roots among $n$ nodes in an event $e$'s subgraph as an $n\times n$ matrix $\mathbf{K}$, with entries
\begin{equation}
\mathbf{K}^{(e)}_{ij} = 
\begin{cases}
	 1, & \text{if the subgraph of node $i$ contains a root created by node $j$ and no forks are observed for $i$ or $j$.} \\ 	
 	0, & \text{otherwise.} 
\end{cases}
\end{equation}

Let $k^{(e)}$ be the sum of all elements in matrix $\mathbf{K}^{(e)}$. The metric $k^{(e)}$ is in the range $[0,1]$.
\begin{equation}
\label{EQ:k}
k^{(e)}= \frac{1}{n^2}  \sum_{i} \sum_{j} \mathbf{K}^{(e)}_{ij}.
\end{equation}
where the numerator is a count of the number of nodes observing each known root in the subgraph. The denominator $n^2$ normalises $k^{(e)}$ to the range [0,1], where $k^{(e)}=1$ if all nodes know all roots. 

\begin{figure}[ht]
\begin{subfigure}{0.45\textwidth} \centering
	\includegraphics[width=\textwidth]{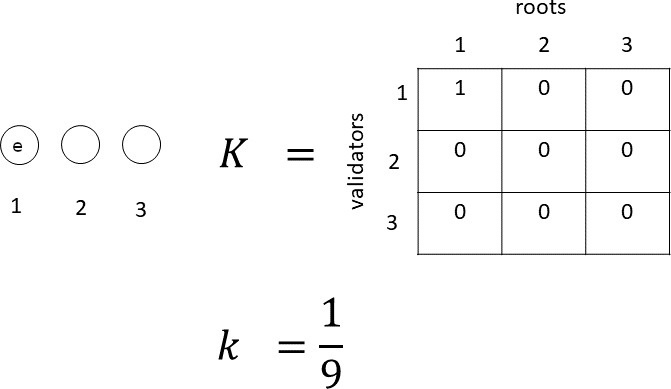}
	\caption{\label{FIG:RootKnowledge1} Root knowledge of event $e$ on the left is $k^{(e)}$= 1/9.}
\end{subfigure}
\hfill
\begin{subfigure}{0.45\textwidth} \centering
	\includegraphics[width=\textwidth]{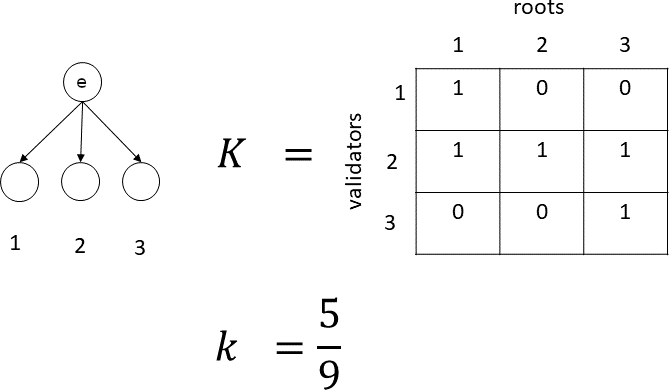}
	\caption{\label{FIG:RootKnowledge2} Root knowledge of event $e$ on the left is $k^{(e)}$= 5/9.}
\end{subfigure}
\caption{Examples of root knowledge matrix $\mathbf{K}^{(e)}$ and root knowledge metric $k^{(e)}$. The network consists of three nodes which index the matrix}
\end{figure}

We define the progress of an event $e$ as a scalar metric of DAG progress, $k^{(e)}$. This scalar metric will enable a comparison the progress of two events.
Figures \ref{FIG:RootKnowledge1} and \ref{FIG:RootKnowledge2} show $\mathbf{K}$ and $k$ examples for simple DAGs. 

A new root will typically be produced before $k^{(e)}$ because producing a new root only requires quorum prior roots to forkless cause the new root.

\subsubsection{Pseudo-code}

The \texttt{rootProgress} function in calculates the root knowledge $k^{(e)}$ of an input event $e$. The function \texttt{rootProgress} does not explicitly construct the $n\times n$ matrix $\mathbf{K}^{(e)}$ in memory. Instead, $k^{(e)}$ can be calculated by looping over the indexes of $\mathbf{K}^{(e)}$ without storing the matrix's elements in memory. The implementation uses function \texttt{ForklessCauseProgress} to calculate each column of $\mathbf{K}^{(e)}$.

\begin{algorithm}
	\caption{\texttt{rootProgress} $k^{(e)}$ calculation}\label{ALG:k}
	\begin{algorithmic}
		\Function{rootProgress}{$e$}		
		\State R $\leftarrow$ roots in $e$'s subgraph
		\State s $\leftarrow$ 0
		\For{r in range(R)}
			\State $c$ = ForklessCauseProgress(e, r)
			\State $s$ = $s$ + sum($c$)
		\EndFor \\
		\Return s
		\EndFunction
	\end{algorithmic}
\end{algorithm}
In Alg.~\ref{ALG:k}, ForklessCauseProgress returns a column of $\mathbf{K}^{(e)}$, corresponding to the number of nodes that have an event whose subgraph contains root, without forks.

\section{Parent Selection and Event Timing}

Based on the metrics, a node can rank candidate events for its new event.
For each QI metric and RK metric, a parent selection method is defined to rank candidate parents into a sorted list. 
Parents are selected from the list if they provide the greatest progress toward producing the next root.

Based on the two metrics, we also define two different event timing strategies. We define new event timing metrics $t$ for a node $v$ based on DAG progress metric (e.g., either QI metric $h^{(e)}$ or RK metric $k^{(e)}$).

We describe the timing metric $t$ based on RK metric $k^{(e)}$ in more details as follows. 
The timing metric $t$ can be defined base on QI metric $h^{(e)}$ in a similar way.

Let $v$ be a node, and let $e_v$ be the latest event of $v$. Within node $v$'s local DAG, for a node $i$, let $e_i$ be the latest event of $i$. 
The new metric $t$ counts the number of nodes $i$, whose highest event has metric $k^{(e_i)}$ that exceeds that of $e_v$.
$t=\sum_i H(k^{(e_i)} - k^{(e_v)}) s_i$,
where $s_i$ is the stake of node $i$ and $H(x)$ is the step function. $H(x)$= 1 \text{if} $x> 0$; 0, \text{otherwise}.

The event timing metric is used to order nodes and is a direct measure of the progress a node has made toward producing a new root and progressing toward the next frame.
Nodes emit an event when they fall below a threshold level in an ordering of nodes, and new emission is to help increase their DAG progress metric.

\section{Simulation Results}
\begin{figure}[htp]\centering
	\begin{subfigure}{0.75\textwidth} \centering
		\includegraphics[width=.9\textwidth]{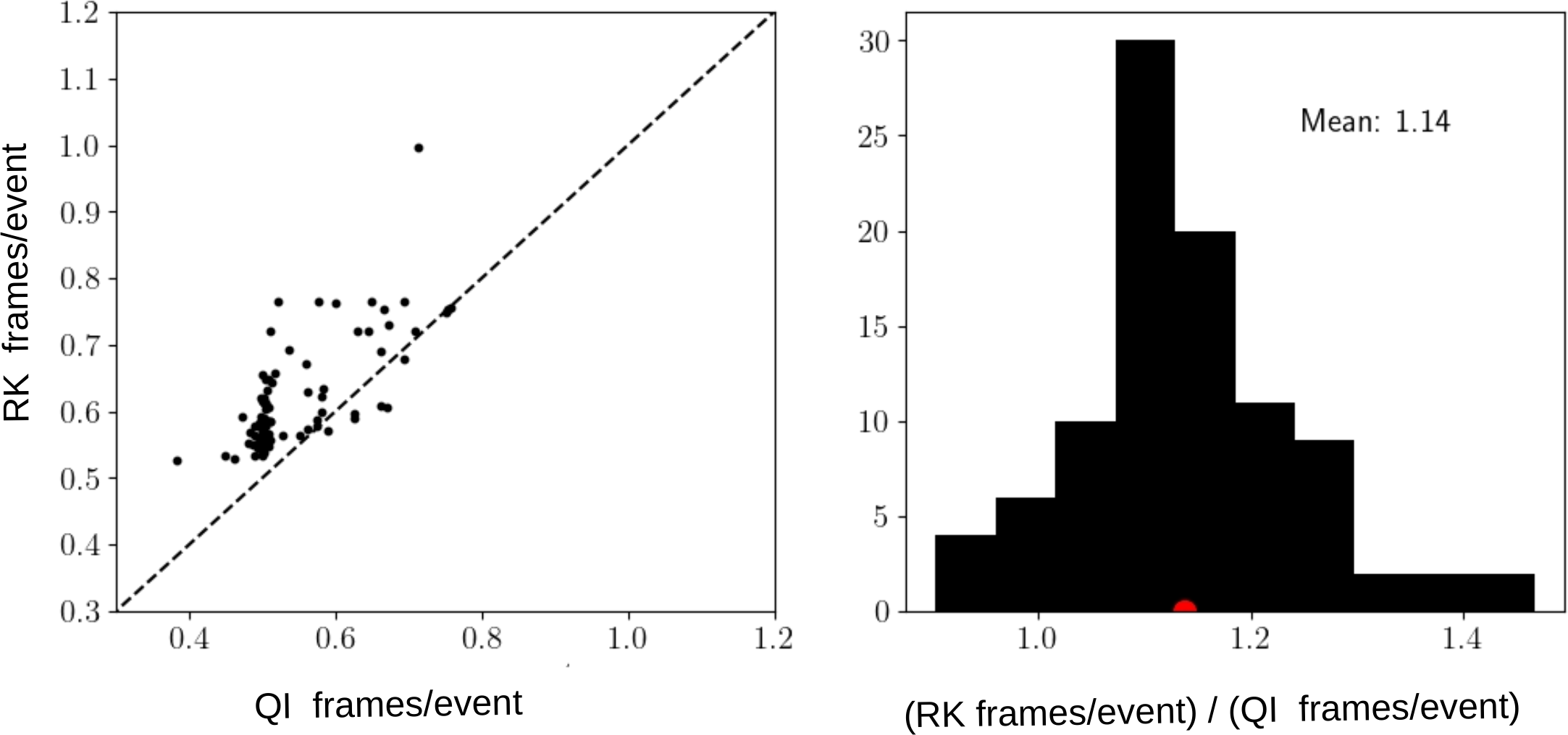}
		\caption{\label{FIG:FramePerEvent} Comparing 'Frames produced Per Event'}
	\end{subfigure}
	\begin{subfigure}{0.75\textwidth} \centering
		\includegraphics[width=0.9\textwidth]{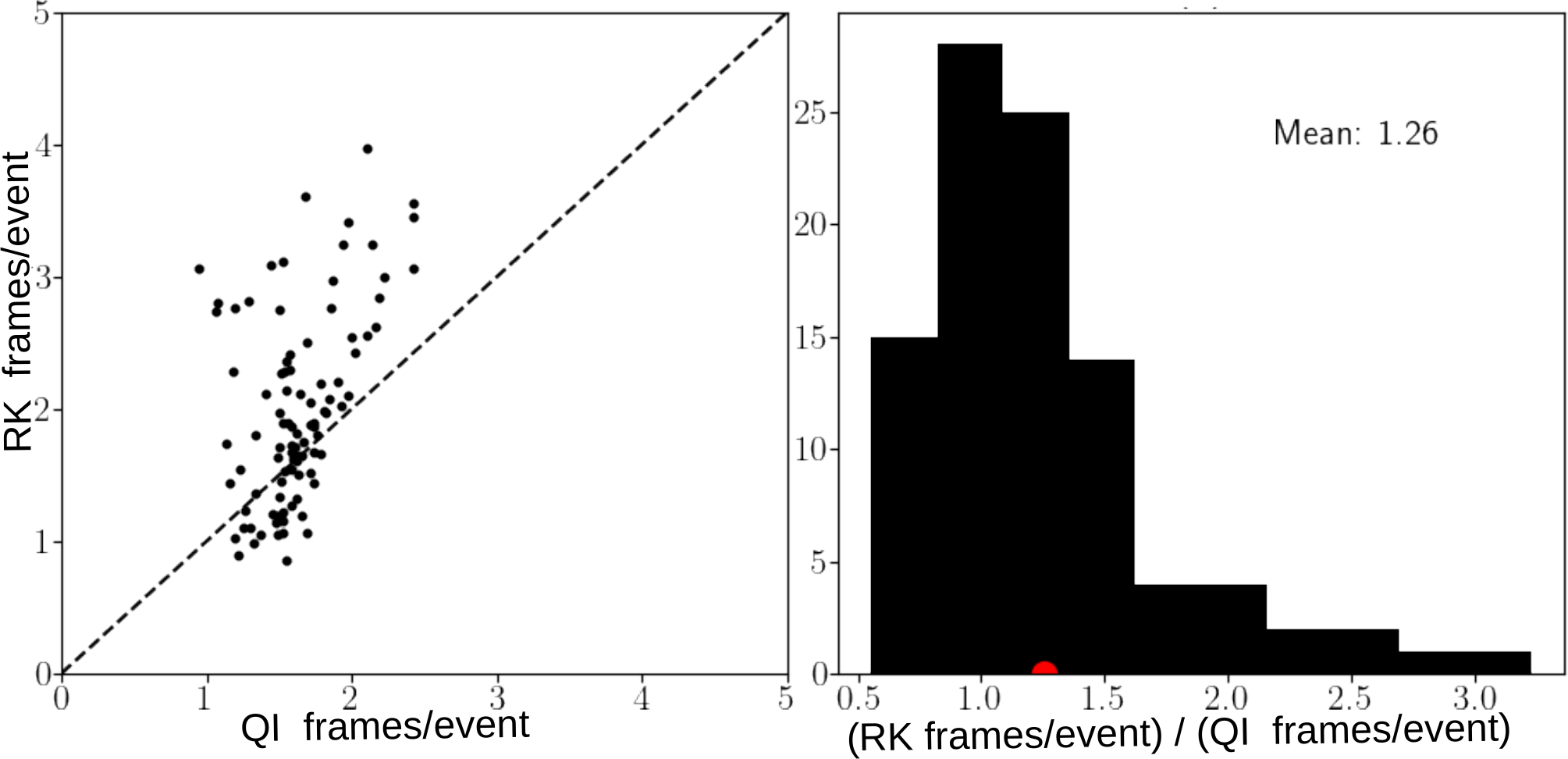}
		\caption{\label{FIG:FramePerSecond} Comparing 'Frames produced Per Second'}
	\end{subfigure}
	\begin{subfigure}{0.75\textwidth} \centering
		\centering
		\includegraphics[width=.9\textwidth]{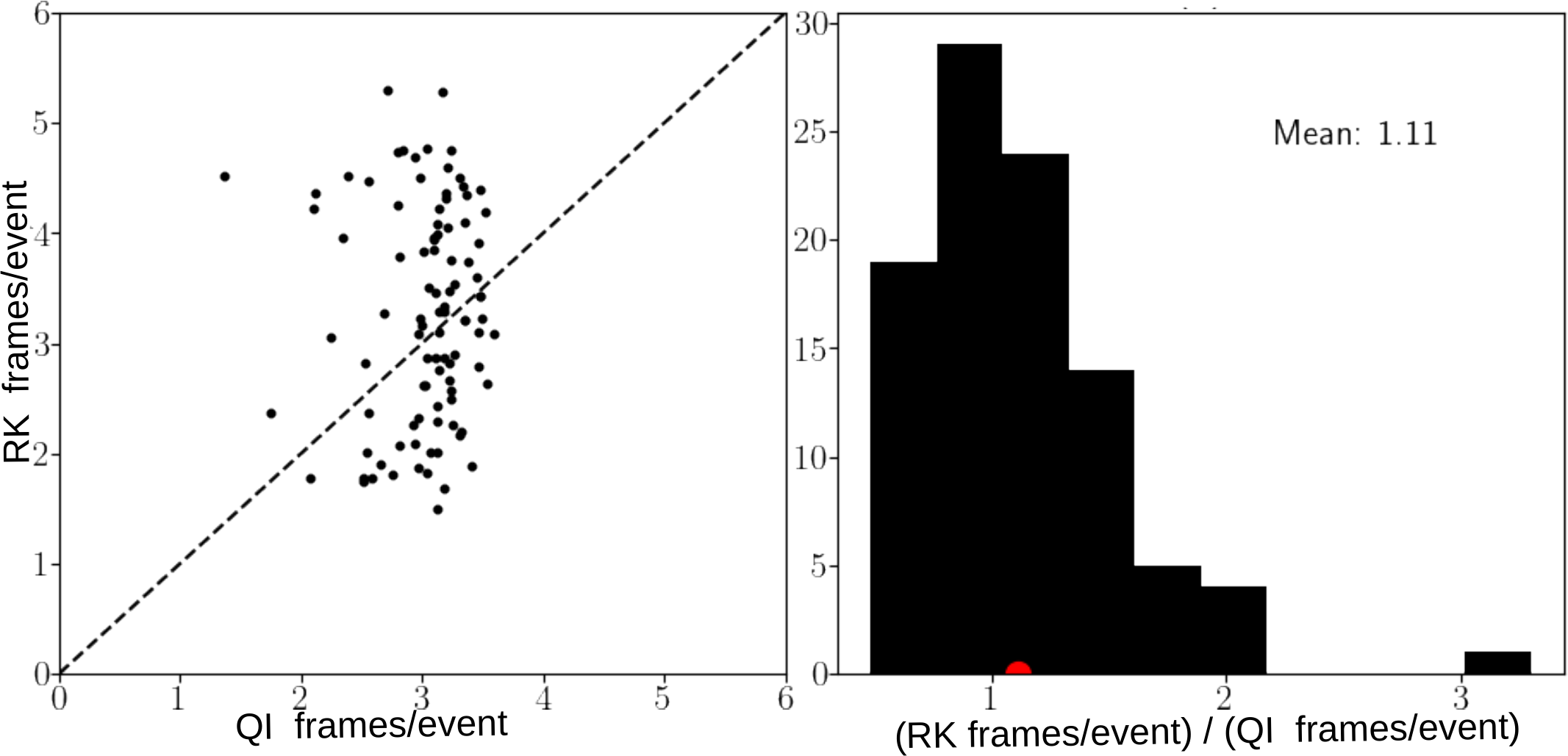}
		\caption{\label{FIG:MeanEventRate} Comparing 'Frame Rate'}
	\end{subfigure}
	\caption{Comparing mean Frame Rate between RK metric and QI metric parent selection and event timing. Larger is better.}
\end{figure}

Figures~\ref{FIG:FramePerEvent}, \ref{FIG:FramePerSecond}, and \ref{FIG:MeanEventRate} compare the performance in simulation between parent selection and event timing methods using QI metric and RK metric.

The experiments were made on a simulation of a network consisting of 30 nodes, with the stake of each node randomly sampled. A dataset of real world internet latencies between cities is used to model latencies between pairs of nodes. It randomly allocated each node to a city in the dataset. Each of the 100 simulations shown are simulations of 100 seconds of network activity.

On average, the new RK metric is more efficient, producing each frame using fewer events compared to the QI metric. The new RK metric produces more frames per second on average compared to QI metric, and thus, the new methods will lower TTF.

\begin{figure}[htb]
	\centering
	\includegraphics[width=0.7\linewidth]{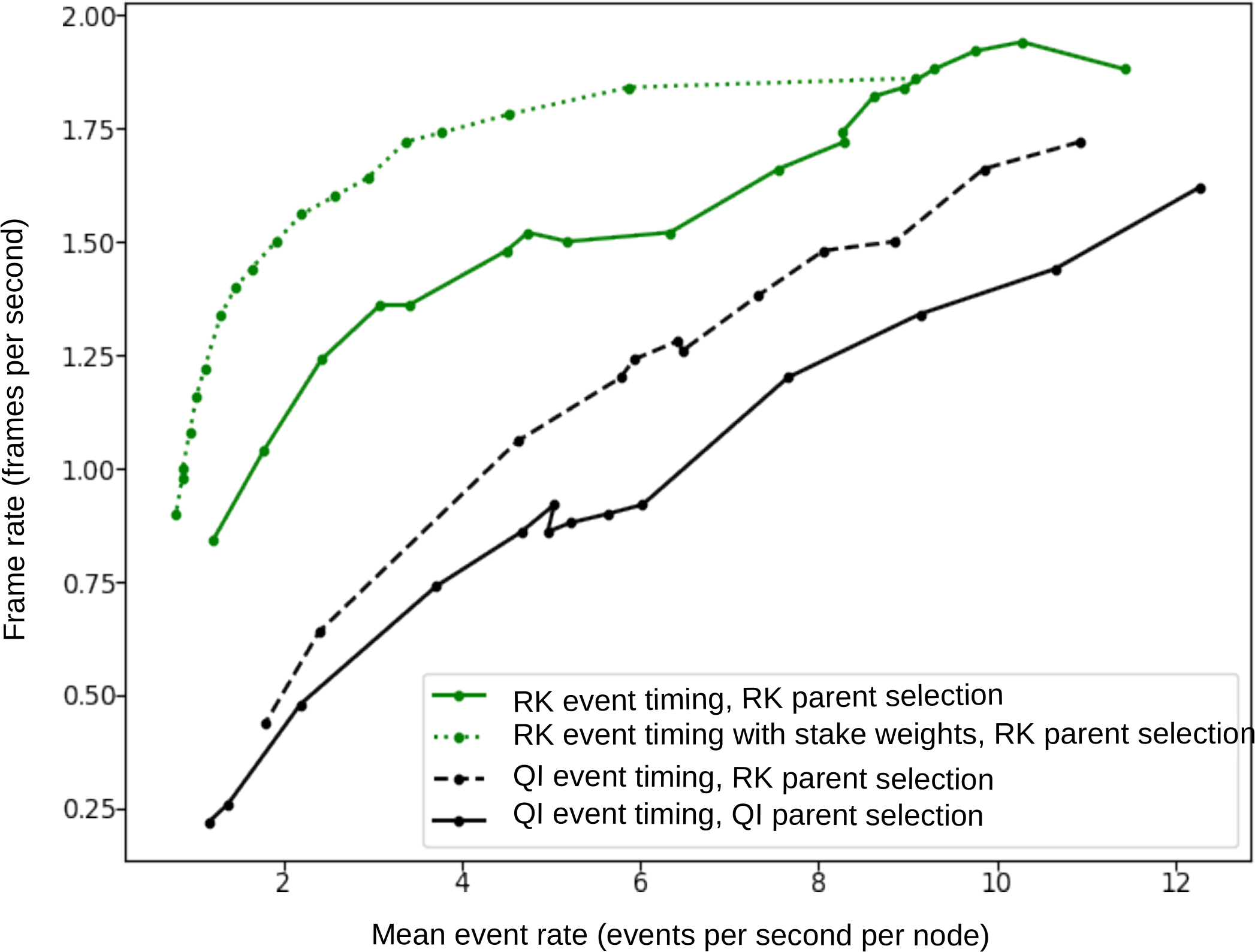}
	\caption{Comparison of frame rate for different combination of event timing and parent selection}
	\label{fig:frameeventratemethodcomparison}
\end{figure}

Figure~\ref{fig:frameeventratemethodcomparison} compares frame rate.
The experiments were made with a simulation of 40 nodes, each event has 3 parents. Latency was set around 100ms.

\section{Conclusion}

In this work, we introduce two new metrics to measure the DAG progress of a new event. 
By comparing between different possible new event blocks it can generate, a node can determine which new event block is the most effective to be made in order to progress toward achieving a new root.

Based on the metrics, we presented new methods for event timing and parent selection. These methods will give several advantages.
First, they can reduce the number of generated events and can avoid generating many excessive events and hence can reduce communication overheads. Second, using these methods, it can achieve faster finality as it can compute roots and Atroposes quicker.	Third, it can reduce events per frame and increase frame rate, as more  events are generated only if they help achieve roots quicker. As such, it improves the number of frames per second (frame rate).

\section{Reference}\label{se:ref}

\renewcommand\refname{\vskip -1cm}
\bibliographystyle{abbrv}
\bibliography{Lachesis}

\end{document}